\documentclass[preprint,12pt,onecolumn]{elsarticle}
\usepackage[dvips]{epsfig}
\usepackage[dvips]{graphicx}
\usepackage{graphicx}
\usepackage{latexsym}
\usepackage{amssymb}
\journal{Computer Standards $\&$ Interfaces}
\begin{document}
\begin{frontmatter}
\title{Mutual Authentication in Self-Organized VANETs}
\author{C\'andido Caballero-Gil, Pino Caballero-Gil, Jezabel Molina-Gil}
\address{Dept. Statistics, Operations Research and Computing.  University of La Laguna. \\38271 La Laguna. Tenerife. Spain. \\E-mail addresses: \{ccabgil, pcaballe, jmmolina\}@ull.es}

\begin{abstract}
The practical deployment of vehicular networks is still a pending issue. In this paper we describe a new  self-organized method of authentication for VANETs, which allows their widespread, fast and secure implementation. Our proposal does not involve any central certification authority because the nodes themselves certify the validity of public keys of the other nodes. On the one hand we propose an algorithm that each node must use to choose the public key certificates for its local store. On the other hand, we also  describe a new node authentication method based on a cryptographic protocol including a zero-knowledge proof that each node must use to convince another node on the possession of certain secret without revealing anything about it, what allows non-encrypted communication during authentication. Thanks to the combination of the aforementioned tools,  the cooperation among vehicles can be used for developing several practical applications of VANETs, such as detection and warning about abnormal traffic conditions. One of the most interesting aspects of our proposal is that it only requires existing devices such as smartphones, because the designed schemes are fully distributed and self-organized. In this work we include an analysis of both an NS-2 simulation and a  real device implementation of the proposed algorithms, which enables us to extract promising conclusions and several possible improvements and open questions for further research.

\end{abstract}

\begin{keyword}
vehicular ad-hoc network, authentication, security, self-organization, wireless communication 
\end{keyword}
\end{frontmatter}

\section{Introduction}

Among the wireless networks that have received more attention of both the research and the industry communities in the last years are Vehicular Ad-hoc NETworks (VANETs). A VANET may be defined as a spontaneous wireless network of vehicles, which allows them to communicate and share information, with the main goal of improving traffic conditions. In particular, communications among vehicles have a tremendous potential to improve road safety, traffic efficiency, and comfort for both drivers and passengers. Therefore, a rapid deployment of VANETs would be very useful to save time and money spent on the road, and to reduce environmental pollution and consumption of fuel reserves. 

Security of communications in VANETs is one of the most important issues to enable their practical deployment because of the variety and severity of possible attacks. 
On the one hand, false traffic warning messages can influence drivers’ decisions, waste drivers time and vehicles fuel, and even lead to traffic accidents. Therefore, VANETs should prevent that attackers can send untruthful information about road conditions such as traffic jams in order to mislead other vehicles. This implies that VANETs should not provide full vehicle anonymity because the possibility to send false messages would compromise their safe practical application. In fact, node authentication is necessary both to guarantee that only trustful vehicles can communicate and to allow law enforcement to track offending vehicles as an aid in investigations about stolen cars or hit-and-run accidents for example. However, on the other hand, VANETs must provide a way to retain privacy in order to avoid that vehicles can be tracked under normal circumstances, because that could provide information about past and current locations of vehicles, what would lead to the lack of drivers privacy and even be misused for crimes such as kidnapping and robbery. In conclusion, since vehicles in VANETs require privacy, it is important to devise a method to authenticate them while maintaining privacy. 

The aforementioned security requisites of VANETs are added to other needs related to efficiency,  such as scalability, cooperation, stability and low communication delay, which should be considered too. All those requirements are more challenging in these networks than in other wireless networks due to their specific characteristics, such as lack of fixed infrastructure and rapidly changing scenarios ranging from rural roads with little traffic to cities or roads with a large number of vehicles. Consequently, communication security can be considered  one of the most challenging research issues that have to be taken into account before carrying out a broad deployment of VANETs. 
In recent years there has been abundant research on vehicular networks, but so far no proposal can be found in the bibliography that imply the feasibility of their secure, broad and rapid deployment of these networks. Nowadays, IEEE standardization efforts are converging towards the definition of the so-called Wireless Access in Vehicular Environment (WAVE) protocol, and  of the draft 802.11p \cite{WAVE} that will be the standard for medium access control in inter-vehicle communications. 

The starting point of this proposal is the conclusion that nowadays it is infeasible to introduce a complete model of VANET according to the classical definition found in the literature and in the 802.11p standards, which include Road Side Units (RSUs) and On Board Units (OBU). The deployment of such a type of VANET would be extremely costly, both for users as they would have to buy new cars or install specific devices (OBUs) in their vehicles, and for the state that would have to deploy a large infrastructure on the roads (RSU) to support VANET services. Thus, in the current global economic situation, such large-scale disbursements are infeasible. Therefore, this paper proposes an alternative self-organized approach to VANETs that does not require any infrastructure and any economic investment neither by users nor by governments. Besides, our proposal could be used as a quick and secure introduction to more comprehensive and standardized VANETs in the future.

This paper is structured as follows. Section 2 reviews some related work. In Section 3, proposals for the generation of public keys, for node characterization and for beacon management are included. 
Section 4 presents a new zero-knowledge authentication protocol, and its analysis through a proof of concept implementation. In Section 5, a new method to choose certificates fpr the local key stores is  described, and  analyzed through simulation. Then, Section 6 includes a brief comparison with other proposals.
Finally, Section 7 presents our conclusions and outlines some topics for future research.

\section{Literature Review}

The main objective of this work is the definition of a simple, scalable and practical design for the immediate deployment of VANETs by exploiting the potential of current smartphones. The proposed scheme is based on the collaboration among users through their mobile devices by providing and obtaining updated information of interest about nearby traffic conditions in order to enable them to choose the best route to their destinations. Our proposal takes into account the gradual deployment of VANETs, because initially they will have neither RSUs nor OBUs, and in fact they will have only a few mobile devices. Since the growth of VANETs will be faster or slower depending on its popularity, acceptance, ease of use and cost, all these features have been prioritized in the design. Thus, scalability, efficiency and minimization requirements have been considered in the scheme proposed here.

In this paper we focus on the first phase of VANET deployment, when the number of devices in the network will be smaller. Once the VANET has spread and the number of vehicles belonging to it has increased, the model should be revised to avoid unnecessary communications that can degrade the network. \cite{petit} includes an analysis of the effect of high vehicle densities in VANET communications under these circumstances. Group-based solutions for authentication are proposed for such  situations in \cite{CaballeroGil2009}, where the specific characteristics of inter-vehicle and vehicle-to-roadside communications are taken into account to define different authentication services. Also a group-based method is proposed in \cite{Zhang12} in particular for 802.11p vehicular networks. 

The practical requirement minimization is a criterion used in several studies focusing on different aspects and applications of VANETs. For example, \cite{Panayappan2007} proposes a notification scheme of free parking lots that does not require any complete infrastructure but only RSUs located in the parking lots. Moreover, \cite{Studer2009} proposes a key management scheme for VANETs, which is used to authenticate messages, identify legitimate vehicles and prevent access to malicious vehicles. However, such a proposal is based on the use of a public key infrastructure, which involves several problems, such as the certification of public keys. On the other hand, with the changing topology of VANETs, it is challenging to sustain  connections for extended periods of time, so broadcasting messages is  the most scalable solution. However, flooding of messages can result in a huge number of collisions in the network and hence in a hard degradation of performance. This particular problem is analyzed in \cite{Hsiao} for the case when signature flooding is used for authentication.

In general, security in VANETs is a critical concern that has been studied by many researchers. For instance, \cite{Raya2005a} uses anonymous certificates to hide the true identities of users, but in that proposal  privacy can still be invaded by tracking  senders until  identities are discovered. The issue of privacy in VANETs is discussed in several papers such as \cite{DDSV} and \cite{Wei2012}.  \cite{Ploil2008} proposes the protection of privacy through the combination of symmetric and asymmetric cryptography. On the other hand, \cite{Wang2008} uses session keys to protect privacy. Finally, \cite{Lin} presents  a privacy-preserving vehicular communications protocol that is based on group signatures, but  its main trouble is that the proposed method cannot deal with the exclusion of compromised vehicles. Another security scheme for  vehicular networks that includes authentication with privacy preservation is \cite{Sha}, where   public key cryptography  is used, and the notion of adaptive privacy  and
 a group-based authentication protocol are proposed. 

There are many references on the issue of node authentication in VANETs that offer different types of self-managed schemes, but using methods that are totally different from the one presented here. For example, \cite{Calandriello2007} proposes an authentication scheme that relies exclusively on pseudonyms, while \cite{Li2008} describes a scheme that combines authentication, key establishment and blind signature techniques. On the other hand, in \cite{Zhang} each RSU maintains an on-the-fly generated group consisting of vehicles that occasionally enter the RSU communication range so that the RSU periodically broadcasts its own certificate and its neighbor RSU certificate to the vehicles within its range. However, verification is not efficient enough due to the length of the signature. With respect to certification of public keys, \cite{Laberteaux2008} presents a method for revoking certificates based on epidemic distribution car-to-car, and \cite{Haas2009} proposes a different mechanism  that needs a central Certification Authority (CA) and certificate revocation lists.

The secure and self-organized approach of VANETs followed in this work is not used in any of the aforementioned papers. In particular, our authentication proposal is focused on enabling the immediate and rapid deployment of VANETs through existing mobile devices.
 
\section{Basic Elements}

The proposed authentication method  is based on a Zero-Knowledge Proof (ZKP), which is a cryptographic protocol that a prover can use to prove possession of a certain piece of information to a verifier without revealing anything about it. 
During the authentication procedure, the prover, denoted $A$, must answer to a number of challenges issued by the verifier, denoted $B$. 
The admission control  included in the  authentication proposal described below uses the general scheme of ZKP defined in \cite{CaballeroGil2006}  based on the graph isomorphism problem, for the particular case of the Hamiltonian Cycle Problem (HCP), which involves the determination of whether a graph contains a cycle that visits each node exactly once. 

Our proposal is based on certificate graphs \cite{Capkun2003}, so that each node $A$ has a private/public key pair and a key store ($KeyStore_A$) including a list of all node certificates that $A$ trusts. 
The set of stored public keys and certificates may be represented as an undirected graph $G = (V, E)$, known as certificate graph, in which each vertex represents both a public key and its owner, and each edge $(A, B)$ symbolizes two public key certificates: of node $A$ signed with the private key of node $B$, and vice versa. A certificate chain is an undirected path in a certificate graph. The subgraph $G_A$ of the certificate graph  $G$ contains exactly the current certificates stored by node $A$ in $KeyStore_A$.

The  following subsections include brief explanations of the generation of  public keys, characterizations of nodes and management of beacons.

\subsection{  Public  Key Generation}
\label{key}

\begin{figure}[htb]
  \centering
\includegraphics[scale=0.57]{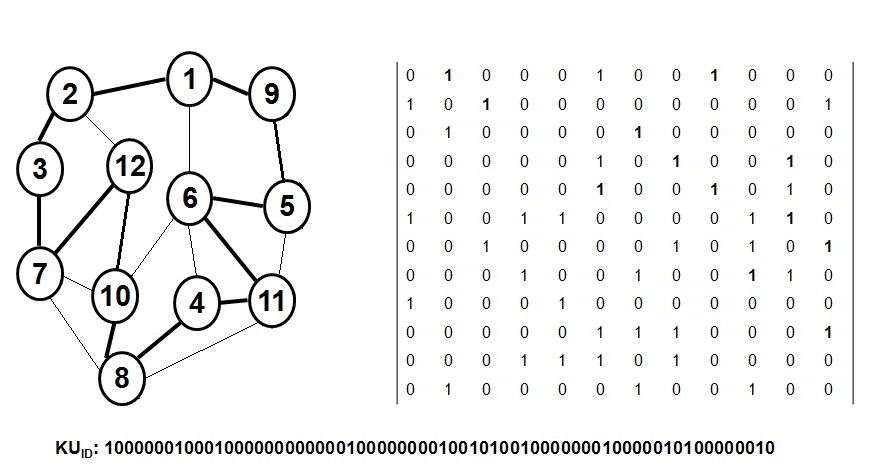}
  \caption{Small Example of HCP based Public Key}
  \label{Fig:Imagen6}
\end{figure}

The   node authentication process described below  is based on the implementation of the ZKP for the HCP. That is the reason why  we use the decimal value of the binary representation for the upper triangular submatrix of the symmetric adjacency matrix containing the elements corresponding to a Hamiltonian cycle in a graph (see Figure \ref{Fig:Imagen6}).
Such a decimal value is used in the proposal as public key in a cryptosystem used by the mobile devices to encrypt messages and to sign public key certificates. 

\begin{figure}[htb]
  \centering
\includegraphics[scale=0.57]{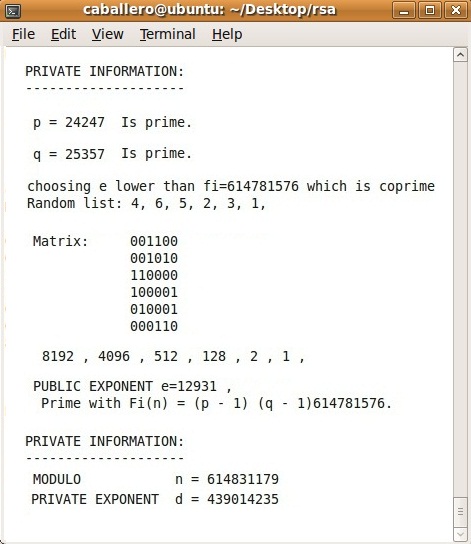}
  \caption{Small Example of RSA based Public Key Generation}
  \label{Fig:RSA}
\end{figure}

In particular,  Figure \ref{Fig:RSA} shows a small example to illustrate the HCP based  public  key generation through an implementation  where the RSA cryptosystem is used. It includes a  trace of the election of a  public key exponent $e$  by using the HCP, as explained above. Thus, after choosing the prime numbers $p$ and $q$, the public exponent $e$ is generated from a random Hamiltonian cycle so that it is lower than and coprime with $(p-1)(q-1)$.
This way to choose  public keys is especially useful because since they correspond to solutions to the HCP in the certificate graph, they can  be directly used in the ZKP based authentication between users for proving their knowledge about them.
 After choosing the public key $e$, the module $n$ and the private exponent $d$ are generated according to the  RSA procedure.

\subsection{Node Characterization}

In general, the  proposal assumes that each node in the network is characterized by the
following elements:

$ID, (KU_{ID}, KR_{ID}), (ID_{i}, KU_{ID_{i}}, Cert(KU_{ID_{i}}))_{ID_{i} \in KeyStore}$ 

corresponding to:

\begin{itemize}
\item A unique IDentifier (denoted ID), obtained as the output of a one-way function on a single value. For example, if the used device is a mobile phone  the value can be its number, while in other cases an email address might be used. The one-way function could be any hash function.
\item A fixed public/private key pair (denoted (KU,KR)) and called identity keys, which are used in an asymmetric cryptosystem such as RSA.
\item A key store containing various IDs and corresponding public keys and certificates, which the node keeps  updated with the algorithm proposed later.
\end{itemize}

 \subsection{Beacon Management}
 \label{Beacons}
The  multicast of beacons containing variable sender identifiers is required both for the active node discovery process and also to avoid vehicle tracking. 
In particular, the variable identifier of each node, which is sent as part of its beacon is the hash of the IDs that are present in its key store at that moment.
In detail, the beacons sent by a node are formed by the following parameters:
\begin{itemize}
\item Frame-Control (FC), which indicates the type of data being sent.
\item Pseudonym (Pseu), which is a temporal identifier of the node. 
\item Timestamp (Time), which allows knowing the specific time when the information was generated. 
\item Pair formed by public key and timestamp (KU, Time) encrypted with the private-key (KR) of the node, which is used by nodes who have already authenticated it when its Pseu changes.

\end{itemize}

\section{Mutual Node Authentication}
\label{Authentication}

Since our proposal is self-organized, the device associated to each network node should be able both to generate its public/private key pair and also to sign the public keys of other nodes that are trustable and want to become part of the network. In order to be able to authenticate its public key to be able to participate in the normal operation of the network, every node must exchange signatures with other legitimate network nodes. The number of necessary signature exchanges will  depend on the current width of the VANET. For practical reasons, at the beginning of the VANET existence, in this proposal two signatures are considered enough to prove that the user is reliable and cannot self-sign certificates to compromise the network security. However, the number of required signatures must grow with the expansion of the VANET.

When two nodes $A$ and  $B$ want to check the validity of each other's public key, they must find a certificate chain between them in the certificate graph that results from merging the subgraphs $G_A$ and $G_B$ corresponding  respectively to $KeyStore_A$ and $KeyStore_B$. 
These $KeyStores$ will have the information necessary to create a certificate chain from $A$ to $B$ as discussed in more detail in Section \ref{KeyStore}.

In particular, the authentication process of the public key of a node $A$ by another  node $B$ and vice versa, is based on a chain of correct and not expired certificates between $A$ and  $B$ in the graph resulting from the union of the two key stores because:

\begin{enumerate}
\item The first certificate in the chain can be verified directly by $A$ (respectively $B$) because it was signed by itself.

\item Each of the other certificates in the chain can be verified by using the public key of the previous certificate in the chain.

\item The last certificate is $B$'s public key (respectively $A$). 
\end{enumerate}

\begin{figure}[htb]
  \centering
\includegraphics[scale=0.60]{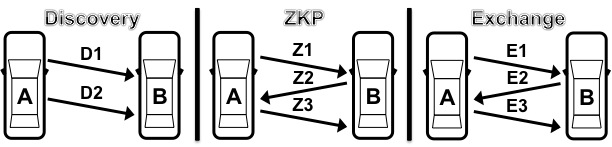}
  \caption{Self-Organized Authentication Protocol}
  \label{FigEN:selfmanagedAuthentication}
\end{figure}

The authentication proposal is composed of three interactive phases in which special packets are sent in order to check the existence of a certificate chain between both nodes and, if it exists, to use it for keys exchange.
Figure \ref{FigEN:selfmanagedAuthentication} shows schematically  these three phases of interaction included in the proposed self-organized protocol for the authentication of the  node $A$ by the node $B$.  In the first phase, both nodes find out whether they are candidates for mutual authentication. In the second phase, they prove that a certificate chain between them exists. Finally, in the last phase, they exchange their keys and key stores.

The three  phases are fully described below. The first phase is the discovering process, which includes part of the  beacons sent by nodes $A$ and $B$, 
containing the hash of the IDs 
stored respectively in $KS_A$ and  $KS_B$.  Within this phase, both nodes find out whether a common public key $x$ exists in both key stores $KS_A$ $\cap$ $KS_B$,
what happens with a high probability thanks to the method of creation of keystores explained in Section \ref{KeyStore}. In such a case, each node generates a graph from such an element so that it is an  HCP solution in such a graph.  Those graphs $G_{A}(x)$ and $G_{B}(x)$ are used in the second phase, which is based on a ZKP, to mutually prove the knowledge of the common key $x$ through open interactive communications that do not reveal its value. In this way, during the last phase, both nodes are sure that they can use the shared key $x$ to exchange their public keys, temporal secret keys and key stores. 
The proposed algorithm not only allows two nodes to authenticate each other, but it also enables them to exchange both fixed public keys, temporal secret keys and current key stores.
In this way, after the authentication process both nodes can exchange messages in a secure way by encrypting them either with their shared temporal secret keys or with each other's public key,  and by signing them with their corresponding private keys, what guarantees both integrity and non-repudiation of messages.

\medskip
\hrule
\smallskip
\textbf{Algorithm} Mutual Authentication Scheme
\hrule
\medskip
\textbf{function} $Authentication\_Scheme()()$ (...)

\vspace{0.2cm}

D1. $A\rightarrow B$ $(B\rightarrow A)$: beacon with $\{ h(ID_{i}):ID_{i} \in KS_{A}\}$ $(\{ h(ID_{i}):ID_{i} \in KS_{B}\})$

D2. $A\rightarrow B$ $(B\rightarrow A)$:  if $\exists x \in KS_{A} \cap KS_{B}$, $G_A(x)$ $(G_{B}(x))$
\vspace{0.2cm}

Z1. $A\rightarrow B$ $(B\rightarrow A)$:  $GI_{A}(x)$  $(GI_{B}(x))$ isomorphic with $G_{A}(x)$  $(G_{B}(x))$

Z2. $B\rightarrow A$ $(A\rightarrow B)$: a binary random challenge $b$ $(a)$

Z3. $A\rightarrow B$ $(B\rightarrow A)$:  

Z3. \quad \quad  \quad \quad If $b=0$ $(a=0)$ $GI_{A}(x)\approx G_{A}(x)$ $(GI_{B}(x)\approx G_{B}(x))$

Z3. \quad \quad  \quad \quad Otherwise a Hamiltonian circuit in $GI_{A}(x)$ $(GI_{B}(x))$

\vspace{0.2cm}

E1. $A\rightarrow B$ $(B\rightarrow A)$: $E_{x}(KU_{A})$ $(E_{x}(KU_{B}))$

E2. $B\rightarrow A$ $(A\rightarrow B)$: $KU_{A}(K_{B})$ $(KU_{B}(K_{A}))$

E3. $A\rightarrow B$ $(B\rightarrow A)$: $E_{KB}(KS_{A})$ $(E_{KA}(KS_{B}))$

\vspace{0.2cm}

\textbf{end function}
\smallskip
\hrule
\medskip

\begin{figure}
  \centering
\includegraphics[scale=0.35]{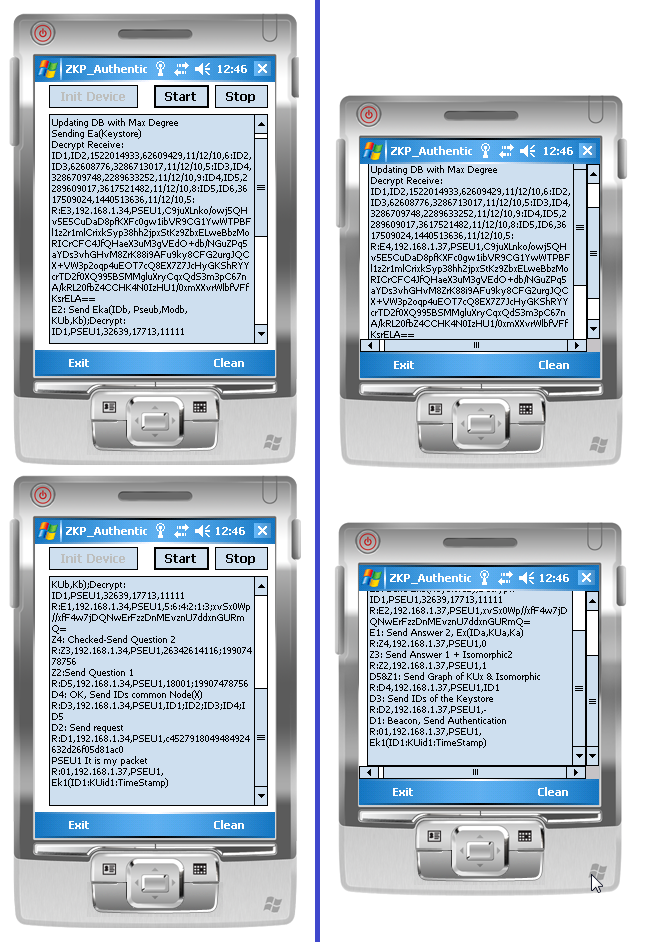}
  \caption{$C\#$ Implementation of the Proposed Authentication Scheme}
  \label{Fig:autenticacion}
\end{figure}

Figure \ref{Fig:autenticacion} shows several screenshots of an implementation of the proposed authentication scheme performed using Microsoft Visual Studio in $C\#$.
A client-server  capable of multiple connections at the same time is implemented in each device. All signals about authentication and beacons are performed with UDP packets. Each client broadcasts beacons periodically  to all connected devices in the network. Each beacon is formed by the following data:

\textit{"01," + thisIpAddr + "," + PSEU + "," + Ek1(ID1,KUid1,TimeStamp)}

Before starting to use the device, the node needs information to communicate with other devices, and in particular a database with three tables is loaded. These tables keep data for a low number of users whose data (certificates and public key) are generated with the generator:

\vspace{0.1cm}
\noindent\textit{certificateStore (idcolumn INT PRIMARY KEY, idA NTEXT, idB NTEXT, certAB BIGINT, certBA BIGINT, date DATETIME); }

\vspace{0.1cm}

\noindent\textit{keyStore (idcolumn INT PRIMARY KEY, idA NTEXT, PseuA NTEXT, module BIGINT, publicKey BIGINT, secretKey BIGINT, degree INT );}

\vspace{0.1cm}
\noindent\textit{myStore  (idcolumn INT PRIMARY KEY, idA NTEXT, PseuA NTEXT, modulo BIGINT, publicKey BIGINT, privateKey BIGINT, secretKey BIGINT, degree INT );}
\vspace{0.1cm}

Incoming connections are managed on the server so that when one is received, the server checks the identity of the node who sent the packet. After that, it checks whether the node is already authenticated in the network, and if not, the authentication protocol begins. 

\section{Key Store Update}
\label{KeyStore}

In the proposal presented in this paper, it is required that each node has its own key store to authenticate other nodes. Since in VANETs the number of users could become huge, here we propose a scheme for storing public key certificates, which exploits the theory of six degrees of separation. Thanks to this property, it is not necessary for each user to store the certificates of all previously authenticated nodes. Instead, it stores only the minimum necessary number of certificates that by merging its store with each other’s, the probability of finding at least a certificate chain in the graph that results from the merge will be high.

Therefore, the optimal update key stores is an important part of the proposal because through it, it is possible to limit the number of stored keys below a value here denoted $lim$. This value is generally smaller than the number of users that form the network, and is equal to the minimum quantity that allows any node to connect to any other node in the network.

In order to maximize the likelihood that any node is capable of authenticating to any other node, while limiting the maximum size of the key stores, different algorithms for updating the key stores may be used. This paper describes a possible algorithm for this. To update its key store, each node chooses the corresponding public key certificates of nodes that have either received or issued more valid certificates, what is represented by the degrees of the vertices in the corresponding certificate graph. This maximizes the probability of intersection between key stores, which is necessary for the proposed authentication process.

\medskip
\hrule
\smallskip
\textbf{Algorithm}  Key Store Update 
\hrule
\medskip
01:\textbf{function} Update\_KeyStore() (...)

02: Initialize data structures;

03: Union:= $KS_A \cup KS_B$; 

04: $KS_B=\{B\}$

05:\textbf{for} each $i \in KS_B$ 

06: \quad \textbf{for} each $j \notin KS_B : (i,j) \in Union$

07: \quad \quad  \textbf{if} ((degree $(j)= max(degree(neighbor of i in Union))$)

    \quad \quad  \quad \quad $\&\& (cardinal(KS_B)< lim))$

08: \quad \quad \quad $(i,j) \in (KS_B)$

09: \quad \quad \textbf{end if}

09: \quad \quad \textbf{end for}

09: \quad \textbf{end for}

09: \textbf{end for}

10: \textbf{end function}

\smallskip
\hrule
\medskip

An implementation of the proposal has been made with the Network Simulator tool NS-2.
In the performed simulation, an initial wireless network where the nodes are located randomly provides the first certificate graph. Each node keeps in its local key store the certificates of nodes at distance 1. Then, nodes begin to move randomly, and when two nodes are within distance 1, they check whether they can trust each other and initiate an exchange of their key stores for their update. New nodes can enter the network by inserting the corresponding new certificates in the certificate graph. Moreover, any node whose certificate is not renewed, is automatically excluded from the certificate graph.

After performing 25 simulations for 15, 20, 30 and 60 nodes, the average results of executions with different types of networks show that the performance can be considered generally acceptable.
According to the simulations we can conclude that the scheme is affected by the mobility of the nodes due to the fact that an increase in their mobility leads to an increase in the speed of growth and balance of their key stores.
Therefore, this is a compelling argument to consider in vehicular networks because they are highly mobile networks. 

\vspace {0.7cm}

\begin{tabular}{|c|c|c|c|c|}
\hline
& {\small 15 nodes} & {\small 20 nodes} & {\small 30 nodes} & {\small 60 nodes} \\
\hline

{\scriptsize Total Connections} &  {\small 966,9}& {\small 1011,52}&{\small 2764,7} & {\small 5309,0}\\

{\scriptsize Successful Connections} & {\small 909,7 }& {\small 985,4 }& {\small 2749,8 }& {\small 5216,5} \\

{\scriptsize Failed Connections}&  {\small 57,15}& {\small 26,12 } & {\small 14,92}& {\small 92,5}\\

{\scriptsize  Added Information} & {\small 102,28 }& {\small 56,4} & {\small 80,51 }& {\small 191,9 }\\

{\scriptsize Key Store Updates} & {\small 628,7}& {\small 420,2 }& {\small 690,24} & {\small 1475,9 }\\

 \hline
\end{tabular}
\label{tab:Trazanodos}

\section{Comparison with Other Proposals}

The following table shows some of the main parameters of our proposal, compared with the ones corresponding to other similar and relevant works. 

\begin{tabular}{|c|c|c|c|c|c|}
\hline
& {\small Wei } & {\small Plobl  } & {\small Li } & {\small Capkun  } &  {\small Caballero } \\
& {\small  \cite{Wei2012}  } & {\small   \cite{Ploil2008}} & {\small \cite{Li2008}  } & {\small   \cite{Capkun2003}} & {\small  } \\
\hline

{\scriptsize Self-Organized} &  {\small yes}& {\small no} &  {\small yes}&{\small yes} & {\small yes}\\
\hline
{\scriptsize Authentication } & {\small no  }& {\small yes }& {\small yes }& {\small yes}& {\small yes} \\
\hline
{\scriptsize Secure}&  {\small yes}& {\small yes } & {\small yes}& {\small yes}& {\small yes}\\
\hline
{\scriptsize  Efficient} & {\small yes }& {\small yes} & {\small yes }& {\small yes }& {\small yes }\\
\hline
{\scriptsize Protect Privacy} & {\small yes}& {\small yes }& {\small yes} & {\small yes }& {\small yes }\\
\hline
{\scriptsize Protect } & {\small yes, with }& {\small not  }& {\small yes} & {\small not  }& {\small yes, with   }\\
{\scriptsize  Anonymity} & {\small  pseudonyms}& {\small  addressed }& {\small } &  {\small  addressed }&{\small   pseudonyms }\\
\hline
{\scriptsize Integrity} & {\small yes}& {\small yes }& {\small yes} & {\small yes }& {\small yes }\\
\hline
{\scriptsize Central CA } & {\small not addressed}& {\small yes }& {\small yes} & {\small no }& {\small no}\\
\hline
{\scriptsize Need } & {\small  not }& {\small not  }& {\small yes} & {\small not  }& {\small not  }\\
{\scriptsize  RSU} & {\small   addressed}& {\small  addressed }& {\small } & {\small  necessary }& {\small necessary  }\\
\hline

{\scriptsize  Simulation} & {\small yes}& {\small no }& {\small no} & {\small yes}& {\small yes}\\
\hline
{\scriptsize Real Device } & {\small no}& {\small no }& {\small no} & {\small no} & {\small yes }\\
{\scriptsize  Implementation} & {\small }& {\small  }& {\small }& {\small } & {\small  }\\
 \hline
\end{tabular}

\vspace {0.2cm}

All the analyzed schemes include security, consider efficiency, and protect user privacy and data integrity. Only one of them (Plobl scheme) is not a self-organized proposal. Regarding authentication, Wei scheme is the only one that does not include any phase for  authentication of users, but of messages. Anonymity is another item that is dealt by most authors, except Plobl and Capkun. Both in Wei and Caballero schemes, the solution for anonymity protection is based on pseudonyms. The issue of public key certification is not  even analyzed in Wei scheme, is solved through a central CA in Plobl and Li schemes, and through a CA distributed among nodes in the two last schemes. Note that reputation and revocation are only discussed in self-organized schemes because they are not an issue in centralized schemes. Indeed,  reputation and revocation are useful tools in distributed schemes to control the behavior of users, and to isolate them if they are impacting negatively in the VANET.
In terms of required infrastructure, it is noteworthy that the two first works do not talk about any interaction between RSUs and OBUs, while Li scheme requires RSUs, and the two last ones avoid RSUs without affecting the operation of the VANET.
With respect to simulation, three of the works  include data about NS-2 simulations, but none of them, apart of this proposal, provide any evidence on real device implementations.

\section{Conclusions}
\label{Conclusion}
In this paper, we have discussed the possibility of developing a self-managed VANET that does not require the deployment of any infrastructure on the road or any special equipment inside vehicles, allowing a gradual introduction of VANETs without any financial investment. To make it possible, the proposed algorithms have been designed so that they can be implemented in existing devices such as smartphones. The main contributions of this work are: a self-managed mutual authentication protocol between nodes based on zero knowledge proofs and certificate graphs, a discovering scheme based of variable pseudonyms to protect privacy and prevent potential tracking, and an algorithm to update local key stores that maximizes the probability of possible communication between any pair of nodes. Our approach allows the use of a hybrid approach that combines secret key cryptography and public key cryptography, what can be used both to optimize resources and to encourage cooperation avoiding the passive behavior of nodes. All proposed algorithms have been simulated with the NS-2 simulator and implemented with Microsoft Visual Studio in $C\#$ on mobile phones. The results obtained in both cases show a high level of performance. Among the open problems to be faced in the near future we can mention the study of specific applications and practical limitations of the proposed schemes for mutual authentication and  key  store update, and their large-scale implementation in real environments. 

\section*{Acknowledgment}
\label{Acknowledgment}

Research supported by the Ministerio de Ciencia e Innovaci\'on and the European FEDER Fund under Project TIN2011-25452 and FPI scholarship BES-2009-016774, and by the ACIISI under FPI scholarship BOC  60.


\begin{thebibliography}{4}


\bibitem{CaballeroGil2009}
P. Caballero-Gil,  C. Caballero-Gil, J. Molina-Gil and C.  Hern\'andez-Goya,
  ''Flexible Authentication in Vehicular Ad Hoc Networks'', Proceedings of the 
 15th Asia-Pacific Conf. Communications, 2009,
  pp. 576--879.

\bibitem{CaballeroGil2006}
P.  Caballero-Gil and C. Hern\'{a}ndez-Goya, ''Zero-Knowledge Hierarchical
  Authentication in MANETs'',
IEICE Transactions on Information and Systems E89-D(3), 2006, pp. 1288--1289.

\bibitem{Calandriello2007}

G.  Calandriello, P.  Papadimitratos,  J.P. Hubaux and  A. Lioy,  ''Efficient and
  Robust Pseudonymous Authentication in VANET'',
 Proceedings of the  4th ACM international workshop on
  Vehicular Ad Hoc networks, ACM, New York, NY, USA, 2007, pp. 19--28. 

\bibitem{Capkun2003}
 S.  Capkun, L.  Buttyan and J.P. Hubaux,  ''Self-organized Public-Key Management
  for Mobile Ad Hoc Networks'',
IEEE Transactions on Mobile Computing 2(1), 2003, pp. 52--64.

\bibitem{DDSV} V. Daza, J. Domingo-Ferrer, F. Sebe and A. Viejo, ''Trust worthy privacy-preserving car-generated announcements in vehicular ad hoc networks'', IEEE Transactions on Vehicular Technology 58(4), 2009, pp. 1876--1886.

\bibitem{Dornbush}
S.  Dornbush and A.  Joshi, ''StreetSmart Traffic: Discovering and Disseminating
  Automobile Congestion Using VANET's'',
Proceedings of the Spring Vehicular Technology Conference, 2007, pp. 11--15. 

\bibitem{Haas2009}
J.J.  Haas, Y.C.  Hu and K.P. Laberteaux, ''Design and Analysis of a Lightweight
  Certificate Revocation Mechanism for VANET'', Proceedings of the Sixth ACM international workshop on
  VehiculAr InterNETworking, ACM, New York, NY, USA, 2009, pp. 89--98. 

\bibitem{Hsiao} H.C. Hsiao, A. Perring F. Bai B. Bellur A.S tuder, C. Chen  and I. Aravind,'' Flooding-resilient broadcast authentication for vanets'', Proceedings of the 17th annual international conference on Mobile computing and networking, 2011, pp. 193--204.

\bibitem{Laberteaux2008}
K.P. Laberteaux, J.J.  Haas and Y.C. Hu, ''Security Certificate Revocation List
  Distribution for VANET'',
Proceedings of the Fifth ACM international workshop on
  VehiculAr Inter-NETworking, ACM, New York, NY, USA, 2008, pp. 88--89. 

\bibitem{Li2008}
C.T.  Li, M.S.   Hwang and Y.P.  Chu, ''A Secure and Efficient Communication Scheme with Authenticated Key Establishment and Privacy Preserving for Vehicular Ad Hoc Networks'',
Computer Communications 31(12), 2008, pp. 2803--2814.

\bibitem{Lin} X. Lin, X. Sun, P. H. Ho and X. Shen, ''GSIS: A secure and privacy preserving protocol for vehicular communications'', IEEE Transaction on Vehicular Technology, 56(6), 2007, pp. 3442--3456.

\bibitem{Panayappan2007}
R.  Panayappan, J.M.  Trivedi, A.  Studer and A. Perrig, ''VANET-Based Approach
  for Parking Space Availability'',
Proceedings of the Fourth ACM international workshop on Vehicular Ad Hoc Networks,  ACM, New York, NY, USA, 2007, pp. 75--76.
 
\bibitem{petit} J.  Petit and Z. Mammeri, ''Analysis of authentication overhead in vehicular networks'', Proceedings of the Third Joint IFIP Wireless and Mobile Networking Conference, 2010. 

\bibitem{Ploil2008}
K. Plobl and  H. Federrath, ''A Privacy Aware and Efficient Security Infrastructure for Vehicular Ad Hoc Networks'',
Computer Standards \& Interfaces 30(6), 2008, pp. 390--397. 

\bibitem{Raya2005a}
M. Raya and J.P.  Hubaux, ''The Security of Vehicular Ad Hoc Networks'', 
Proceedings of the 3rd ACM workshop on Security of Ad Hoc and
  Sensor Networks,  2005, pp. 11--21.
  
\bibitem{Sha} K. Sha, Y. Xi, W. Shi, L. Schwiebert, and T. Zhang, ''Adaptive privacy- preserving authentication in vehicular networks'', Proceedings of the International Conference on Communications and Networking in China, Beijing, China, October 2006, pp. 1--8.  

\bibitem{Studer2009}
A.  Studer, E. Shi,  F. Bai and  A. Perrig,  ''TACKing together Efficient Authentication, Revocation, and Privacy in VANETs'',
 Proceedings of the 6th Annual IEEE communications society conference on Sensor, Mesh and Ad Hoc Communications and Networks,  2009, pp.
  484--492.

\bibitem{Wang2008}
N.W.  Wang,  Y.M. Huang and W.M. Chen,  ''A Novel Secure Communication Scheme in
  Vehicular Ad Hoc Networks''
Computer Communications 31 (12),   2008, pp. 2827--2837.
 
 \bibitem{WAVE} Wireless LAN Medium Access Control (MAC) and Physical Layer (PHY) Specifications: Wireless Access in Vehicular Environment IEEE Std 802.11p/D7.0, 2009.
 
\bibitem{Wei2012}
Y.C. Wei and Y.M. Chen, ''Efficient Self-Organized Trust Management in Location Privacy Enhanced VANETs'', 
Proceedings of the  13th International Workshop on Information Security Applications, 2012.
 
\bibitem{Zhang} L. Zhang, Q. Wu and A. Solanas, ''A scalable robust authentication protocol for secure vehicular communication'', IEEE Transactions on Vehicular Technology 59(4), 2010, pp. 1606--1617.

\bibitem{Zhang12} Z. Zhang, A. Boukerch and H. Ramadan, ''Design of a lightweight authentication scheme for IEEE 802.11p vehicular networks'', Ad Hoc Networks
 10(2), March 2012, pp. 243--252.


\end{thebibliography}
\end{document}